# Precision measurements of Linear Scattering Density using Muon Tomography

**E. Åström,**[g] **G. Bonomi,**[b] **I. Calliari,**[d1] **P. Calvini,**[c] **P. Checchia,**[f,1] **A. Donzella,**[b] **E. Faraci,**[a] **F. Forsberg,**[g,e] **F. Gonella,**[f] **X. Hu,**[h] **J. Klinger,**[f] **L. Sundqvist Ödqvist,**[h] **D. Pagano,**[b] **A. Rigoni ,**[d2] **E. Ramous ,**[d1] **M. Urbani,**[d2] **S. Vanini,**[d2] **A. Zenoni,**[b] **G. Zumerle.**[d2]

[a]*Centro Sviluppo Materiali SPA, Rome Italy,*
[b]*Department of Mechanical and Industrial Engineering, University of Brescia, Brescia, Italy,*
[c]*Department of Physics, University of Genova and Sezione INFN di Genova, Genova, Italy,*
[d1]*Department of Industrial Engineering, University of Padova, Padova, Italy,*
[d2]*Department of Physics and Astronomy, University of Padova, Padova, Italy,*
[e]*Division of Fluid and Experimental Mechanics, LuleåUniversity of Technology, Luleå, Sweden,*
[f]*INFN Sezione di Padova, Padova, Italy,*
[g]*LKAB R&D, Luleå, Sweden,*
[h]*Swerea MEFOS AB, Luleå, Sweden.*

E-mail: paolo.checchia@pd.infn.it

Abstract: We demonstrate that muon tomography can be used to precisely measure the properties of various materials. The materials which have been considered have been extracted from an experimental blast furnace, including carbon (coke) and iron oxides, for which measurements of the linear scattering density relative to the mass density have been performed with an absolute precision of 10%. We report the procedures that are used in order to obtain such precision, and a discussion is presented to address the expected performance of the technique when applied to heavier materials. The results we obtain do not depend on the specific type of material considered and therefore they can be extended to any application.

---

[1]Corresponding author.

# Contents

## 1 Introduction

Muon radiographic methods facilitate the exploration of inaccessible volumes by utilising the highly penetrative property of muons, which are produced in the Earth's upper atmosphere by cosmic rays [1, 2]. The technique of muon scattering tomography (MST), which was first proposed in 2003 [3] as an extension of traditional muon radiography methods, is based on the measurement of the multiple Coulomb scattering of muons crossing the target volume. The method evaluates the spatial distribution, within the target volume, of the Linear Scattering Density (LSD), $\lambda$, which is defined as the inverse of the radiation length $X_0$ [4] in units of rad$^2$/m.

The precise measurement of the LSD of a material is challenging for several reasons, particularly relating to the maximum likelihood expectation maximization (MLEM) algorithmic implementation of MST [5]. For example, in a target volume subdivided into many voxels, the measurement of LSD in each voxel is not completely independent from any other voxel in the set, which leads to an intrinsic "leakage" of LSD from voxels of high LSD to surrounding voxels. Specifically, the scattering variable assigned to every voxel associated with a given muon's path is correlated with one another, which leads to distortions in the prevailing (generally vertical) direction of muons.

For a given material with mass density $\rho$ and average atomic number $Z$, the LSD of the material is approximately proportional to the product $\rho Z$. For this reason, most of the applications which have been proposed for MST relate to the detection of high-$\rho$, high-$Z$ objects embedded amongst low- or medium-$\rho$ and low-$Z$ materials [6–8]. Precision measurements of LSD are of minor importance for such applications, as they require a high degree of contrast between the dense objects and the surrounding material [9].

There are cases where traditional invasive imaging strategies cannot be employed, and so precision measurements of the LSD using MST can provide important information on otherwise inaccessible internal structures. Furthermore, due to the dependency of $\lambda$ on $Z$, MST can be used to discriminate the high-$Z$ content of materials. High precision LSD measurements are therefore an attractive feature for applications such as the monitoring of blast furnaces [10], which aim to improve the energy efficiency of the production process.

In this paper, the principles of tomographic image reconstruction with multiple Coulomb scattering are outlined in Section 2. We then present a procedure that has been developed by performing a campaign of LSD measurements of material samples provided from the LKAB experimental blast furnace [11] (EBF), as described in Section 3. The muon scattering data, which are used for the analysis of the samples, are collected with the muon tomography station prototype located at the INFN LNL laboratory, as described in Section 4. We obtain a reliable and precise measurement of the LSD using our software implementation of the MLEM muon tomography algorithm, as described in Section 4, and results are given in Section 5.

This study shows that is possible to obtain LSD measurements with a precision of about 10%. Our long campaign of precision-dedicated measurements of LSD represents a crucial milestone in the maturity of MST as an imaging technology in several fields including blast furnace applications.

## 2 Tomographic image reconstruction with multiple Coulomb scattering

This Section will outline the tomographic image reconstruction procedure that is used in this study. Several details of the MLEM algorithm and the particular implementation presented in this paper are described in Ref. [5] and Ref [9] respectively. The image reconstruction is based on the information inferred from muons which pass through two detectors that are positioned on either side of a target volume. Whilst traversing the material in the volume, muons will undergo two types of interactions:

- the ionization and excitation of atoms in the material, which results in muon energy loss; and

- muon-nucleon interactions, including multiple Coulomb scattering (MCS), which cause a muon's path to deviate.

From the perspective of performing muon tomography, the effect of the second type of interactions is that muons emerge from a target volume with a different direction, with respect to the direction that was observed prior to entering the volume. The scattering angle, defined as the angle between the "ingoing" and "outgoing" muon direction vectors, is usually projected on a plane containing the incoming muon trajectory. Since the scattering process is the result of a random superposition of a very large number of small individual scatterings, the projected angle is, to a good approximation, distributed in a Gaussian manner for muons of a fixed momentum $p$ and path length $\ell$. The mean of the Gaussian distribution is zero and the variance $\sigma^2$ is related to the muon momentum, path length and the LSD of the target material by the following approximate relation:

$$\sigma^2 \approx b^2 \frac{1}{p^2} \ell \lambda \tag{2.1}$$

where $b = 0.0136$ GeV/c and the LSD $\lambda$ is measured in units rad$^2$/m, when $\ell$ is measured in meters. Formally, Eq. 2.1 is only valid for high energy muons and low thickness targets. According to [9], we also define the muon displacement as the distance between the exit point of the muon from the crossed material and the straight line trajectory of the incoming muon. The displacement also, once projected on a plane containing the incoming muon trajectory, has a Gaussian distribution. It is partially correlated with the scattering angle projected on the same plane.

Muon scattering tomography is based on the measurement of the scattering angles and displacements of a large number of muons crossing the target volume. The MLEM algorithm produces spatially distributed values of LSD by acting on a set of voxels, which are obtained by dividing the target volume into finite volume elements. Each voxel is assumed to contain a single homogeneous material. A schematic representation of the scattering of a single muon is shown in Fig. 1. The $\lambda_j$ value in each voxel $j$ is estimated by maximizing a likelihood function of the measured scattering angles. The contribution to the likelihood due to the scattering of muon $i$ in voxel $j$ is described by the quantity $s_{ij}$ [9], which has the reciprocal dimensions of LSD.

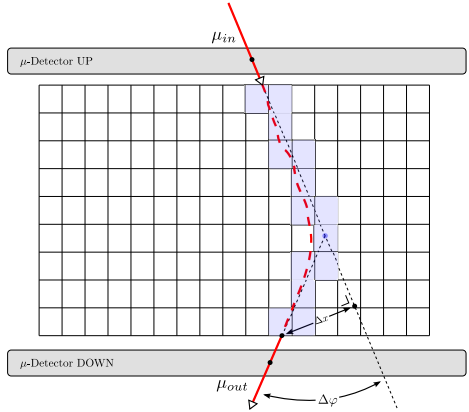

**Figure 1**. A schematic representation of the scattering of a single muon. The direction of the incoming muon lies in the plane of the figure. Both incoming and outgoing directions are shown, together with the scattering angle $\Delta\varphi$ and the displacement $\Delta x$, projected onto the plane of the figure. This figure has been taken from Ref. [9].

The iterative procedure of the MLEM starts from a set of initial values of LSD, $\{\lambda_j^{(n=0)}\}$, and subsequently improves the approximation of LSD based on the distribution of $\{s_{ij}^{(n=0)}\}$. At each step $n$, the expected values of $\{s_{ij}^{(n)}\}$ are updated using the values $\{\lambda_j^{(n)}\}$, and an improved set $\{\lambda_j^{(n+1)}\}$ is obtained by maximizing the likelihood with the values $\{s_{ij}^{(n)}\}$. The correction to each value of LSD at each iteration, $\delta_j^{(n)} = \lambda_j^{(n+1)} - \lambda_j^{(n)}$, depends linearly on the average of the $s_{ij}^{(n)}$ values for all the muons crossing voxel $j$:

$$\delta_j^{(n)} = \left(\lambda_j^{(n)}\right)^2 \frac{1}{m_j} \sum_{i,\ell_{ij}>0} s_{ij}^{(n)}, \tag{2.2}$$

where $\ell_{ij}$ is the path length of the muon $i$ in the voxel $j$ and $m_j$ is the number of muons crossing the $j^{\text{th}}$ voxel.

The MLEM algorithm therefore allows one to approximate the spatial distribution of $\lambda$ within a target volume, which can then be used to analyze the material properties of the volume. The LSD of a material is proportional to the density $\rho$:

$$\lambda = \rho R. \quad (2.3)$$

In case the material is composed of a single element, $R$ is a function of the atomic mass $A$ and of the atomic number $Z$, and it is approximately proportional to $Z$. Consequently we have:

$$\lambda \stackrel{\propto}{\sim} \rho Z, \quad (2.4)$$

as discussed in the introduction. We obtain a precise evaluation of $\lambda$ and $R$ of all elements from Ref. [12].

For objects containing several elements, both in form of chemical compounds or as a non-uniform mixture of different materials, the LSD is still proportional to the bulk density of the materials. The function $R$ is then connected to the $R_i$ functions of the constituent elements by:

$$R = \Sigma_i w_i R_i, \quad (2.5)$$

where $w_i$ is the fraction of the total material mass due to the $i^{\text{th}}$ element. In practice, the quantity $R$ is approximately proportional to the average atomic number of the compounds within a non-uniform material, calculated with respect to the mass fraction of each component. Clearly, $R$ is not sensitive to the presence of air-filled spaces within a target volume as they have a negligible mass.

## 3 Description of material samples

This Section provides a brief description of the material samples which have been probed at the muon tomography station at the INFN Legnaro Laboratories (LNL). The complete set of samples includes specific calibration samples in addition to the samples extracted from the EBF.

The material contained in the upper part of a blast furnace is essentially coke, which is mainly carbon, and pellets of ore, which are composed of various iron oxides. The process of iron production is such that in the central and lower regions of a furnace, one finds coke and a mixture of iron oxide pellets with an increasing degree of reduction. With MST one would expect to measure a range of LSD from low values corresponding to carbon, to high values corresponding to iron. However since the liquid pig iron is concentrated in the lowest region of the furnace, the LSD values which are considered in this study are in the range 1.3 rad$^2$/m (coke) to 14 rad$^2$/m (iron oxides). Several calibration materials have been procured in order to test the response of the MST in presence of calibrations samples with well-established properties. Each calibration sample is composed of a single material and can be considered to be homogeneous on the length scale of the muon tomography spatial resolution. Although the samples also contain small regions of air, this does not affect the measurement of $R$, as discussed in Section 2. The calibration materials that have been analyzed are iron ore pellets with different degree of reduction: a) hematite (no reduction);

b) magnetite (11%-reduction); c) wustite (30%-reduction); d) semi-metallic iron (80%-reduction); plus samples containing: e) coke; f) water.

Each calibration material is stored in a 4.7 liter plastic bucket with average diameter and height of approximately 200 mm and 160 mm, respectively. The effective density of each sample has been evaluated using the mass of the sample and the effective volume occupied by the sample material within the bucket. The measurement of the sample mass is performed with high precision so its error is negligible. The bulk density measurements are therefore assigned an uncertainty of approximately 5% due to the uncertainty on the volume determination. The only exception is for the samples of water, the density of which is precisely known.

Table 1 shows the measurements of the effective bulk densities for the complete set of calibration samples. The calibration samples are referred to by the codes A1→A8. The expected LSD of the calibration materials, to which the corresponding measurements of LSD can be compared, can be calculated using Eq. 2.5 by considering the chemical composition of each material.

| Name | Material | Mass (kg) | Bulk density (kg/dm$^3$) |
| --- | --- | --- | --- |
| A1 | Hematite | 9.959 | 2.34 |
| A2 | Magnetite | 9.400 | 2.12 |
| A3 | Wustite | 9.971 | 2.13 |
| A4 | Semi-metallic iron | 8.119 | 1.83 |
| A5 | Coke | 2.396 | 0.52 |
| A6 | Water | 4.600 | 1.00 |
| A7/8 | Water | 2.400 | 1.00 |

**Table 1**. Mass and bulk density of the calibration samples

The set of samples subsequently analyzed in the LNL muon tomography station provides a partial but significant representation of the distribution of the EBF contents. A portion of the EBF material is extracted during operations from the furnace by means of three probes acting at different heights of the EBF body [11]. The extracted samples are indicated as B1→B3, C1→C3 and D1→D2 for the three probes respectively. A further set of materials is extracted from various positions after extinguishing the EBF. The samplings of this set were excavated from eight different levels in the EBF, with samples extracted from three different positions corresponding to the center, the mid-radius and the wall regions. The naming convention assigns a capital letter (from E to L) from the uppermost to the lowermost level, and a number (from 1 to 3) from the center of the furnace to the wall. The bulk density of the probe and excavation materials were measured using the same methodology as for the calibration materials, with an estimated uncertainty of 5%. The masses of the probe and excavation samples range from about 2.5 kg to about 9.2 kg and consequently the densities varies from 0.56 kg/dm$^3$ to 2.02 kg/dm$^3$. One should note that the measured densities of the EBF samples are within the range of densities of the calibration materials.

## 4 Experimental procedure

This Section describes the data acquisition process with the experimental muon tomography station at LNL, and the subsequent analysis to extract LSD values from the cosmic ray muon data. In Section 4.1 a description of the LNL muon tomography station is given. Section 4.2 provides a brief description of the raw data processing and Sections 4.3 to 4.7 describe the different elements of the data analysis procedure.

### 4.1 Experimental setup at LNL muon tomography station

The samples described in the previous Section were inspected in the muon tomography station prototype at the INFN Legnaro Laboratories. We report the main characteristics of the apparatus, shown in Fig. 2 and described in detail in Ref. [7, 9]. Two 300 cm × 250 cm muon detectors built as spare modules for the CMS experiment [13] (CERN, Geneva), are used for tracking the muons. The two detectors are placed horizontally, with a 160 cm vertical gap, enclosing a volume of 11 m$^3$. The upper muon chamber is set to trigger on events where the muon track points toward the lower chamber. The trigger rate is approximately 350 Hz. The samples are inserted into the inspection volume, as shown in Fig. 2. All data are collected by using the same spatial positions for the measured objects.

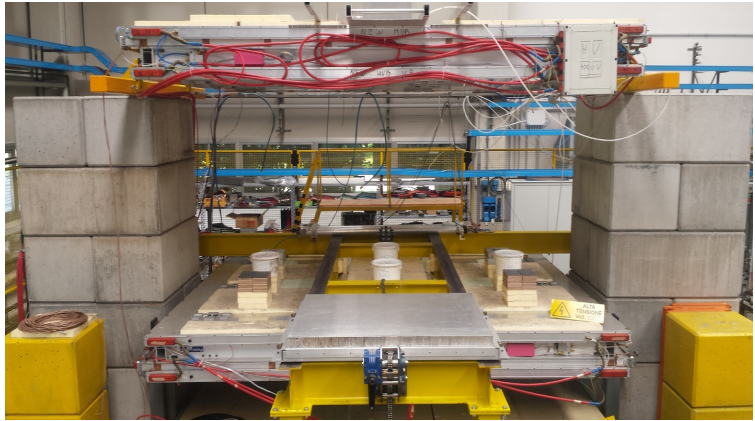

**Figure 2**. The muon tomography station at the INFN Legnaro Laboratories with the samples (white buckets) inserted in the inspection volume. For reference, two mock-up blocks and two iron blocks are placed near the four corners of the lower chamber. A mechanical structure made of iron I-beams (painted yellow) with a mechanical engine (black) is also visible.

### 4.2 Data processing

The raw muon data are processed with the muon tomography software chain. The raw data from the data acquisition system are processed with a pattern recognition algorithm which computes the muon track entering and exiting the inspection volume. The reconstructed tracks feed the LSD reconstruction algorithm, which uses information describing the scattering and displacement of muons due to their respective paths in the inspection volume. Due to the particular construction properties of the muon chambers, which are optimized for the CMS experiment but not optimized for this application, the pattern recognition and track reconstruction efficiency in each chamber is approximately 75%. Since the LSD reconstruction algorithm needs to have the track measured in both chambers, the overall efficiency is about 50%. Therefore, for each approximately two day period of data-taking (which is representative of the total exposure for each sample measurement), the equivalent of approximately one day of data is acquired, with respect to a fully efficient system. The dataset for each run is divided into subsets of 136 minutes, each corresponding to about 1.4 million muons used by the LSD reconstruction algorithm.

For each 136 minute sample, the algorithm produces an image of the target volume, subdividing it into cubic voxels of size $(2.5 \text{ cm})^3$ and estimating the average LSD value in each voxel. Each image is subsequently analyzed to obtain the average LSD value of each sample present in the volume. The final results for each dataset are obtained by averaging the results obtained from each subset.

### 4.3 Noise Reduction

As described in Eq. (2.1), the scattering angle of muons crossing the materials depends on the muon momentum, such that low momentum muons have, on average, larger scattering angles than high momentum muons. Clearly the knowledge of muon momentum would allow to extract the maximal information about the LSD of a sample from CMT data. Suggestions on how to obtain an evaluation of single-muon momentum with an addition of detectors in a muon tomography system are given in recent article [14]. Unfortunately our muon tomography station is unable to determine the momentum of the crossing muons, and hence we must replace Eq. 2.1 with

$$\sigma^2 \approx b^2 \left\langle 1/p^2 \right\rangle \ell \lambda. \tag{4.1}$$

To compute the value of LSD from the measurement of the scattering angle variance we must know the value of $\left\langle 1/p^2 \right\rangle$. Changing this value will rescale the reconstructed LSD values and the issue of the absolute scale will be discussed in Section 4.5. The use of a fixed momentum value leads to the algorithm over-weighting the scatterings of low momentum muons. When this happens, the respective $s_{ij}$ values become so large that they bias the average value of $s_{ij}$ in Eq. 2.2. Consequently, high noise appears in the reconstructed images, as shown in Fig. 3 (left panel), where the tomographic reconstruction of the setup shown in Fig. 2 is presented. The iron blocks, the mock-up blocks, the buckets and the iron structure are clearly visible in the figure, however the image is contaminated by noisy voxels with very high LSD.

In order to reduce the level of noise in the reconstructed images, the following procedure is implemented. Muons with a very low momentum are expected to have, in many voxels, $s_{ij}$ values that are larger than the voxel average. A muon $i$ is therefore removed from the analyzed collection if more than 30% of the voxels crossed by the muon satisfy the relation:

$$s_{ij} > N \left\langle s_j \right\rangle, \tag{4.2}$$

where $\left\langle s_j \right\rangle$ is the average of the $s_{ij}$ values for all muons crossing voxel $j$. The value of the tunable parameter N determines the fraction of muons that are rejected. Rejecting too many muons with large scattering angle could produce a bias in the reconstructed image, because large scattering angles can be due to large values of LSD of crossed voxels, rather than due to low muon momentum. It is

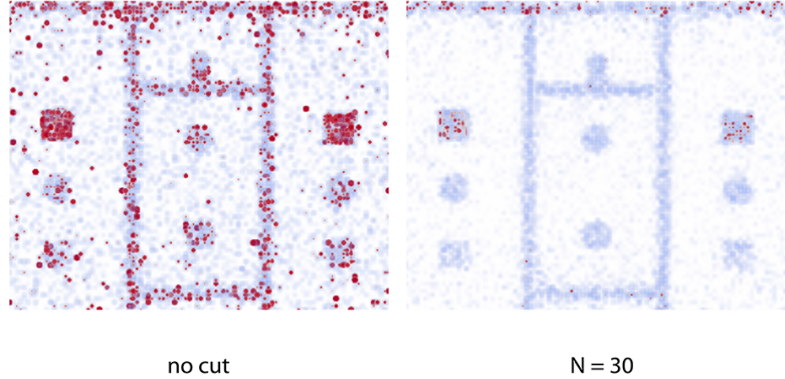

no cut              N = 30

**Figure 3**. Reconstructed images of the setup shown in Fig. 2 without a cut on $s_{ij}$ (left panel), and with a cut on $s_{ij}$ with N = 30 (right panel). The red dots represent voxels in which the LSD value can overflow the upper limit of the colour scale used. See text for details.

therefore important to reject the minimum number of muons compatible with the noise reduction. Various values for this parameter were considered, by calculating the fraction of rejected muons for a given value of N. For example, for N = 10 less than 6% of muons are rejected and the fraction of rejected muons decreases as N increases. For high values of N, for example N = 60, the images display noisy voxels whilst an "intermediate" cut of N = 30 is found to reduce noise well, rejecting only about 2% of muons. The image reconstructed when this cut is applied is shown in Fig. 3 (right panel), which clearly shows noise reduction with respect to Fig. 3 (left panel).

### 4.4 Algorithm Convergence

Image reconstruction with muon tomography is very challenging from the point of view of computing resources, since the number of variables to estimate is very large (of the order of $10^5$ in this case) and the number of measured muons used in the algorithm is also very large (of the order of $10^6$ to $10^7$). Issues relating to RAM can clearly be managed, given sufficient resources. Nevertheless, the algorithm is known to converge slowly. For applications where one searches for high LSD objects amongst volumes with lower values of LSD, sufficiently good discrimination of the materials is reached well before algorithm convergence. For our purposes however, it is required that the level of convergence is sufficiently high in order to minimize residual variations of LSD values between data sets, which would induce large uncertainties in the final result. Since every step of the convergence process is very expensive in terms of computing resources, it is necessary to determine the minimum number of iterations required to achieve a satisfactory convergence level. It has be chosen, for this analysis, to stop the convergence process after 5000 iterations.

### 4.5 Calibration procedure

As discussed in Section 4.3, the absolute scale of the reconstructed LSD values is linearly related to the value of $\langle 1/p^2 \rangle$, a quantity that is difficult to estimate precisely. A practical solution to this problem is to have, in the target volume, objects of known LSD to be used as points of reference. Given two uncalibrated measurements of LSD $\lambda_{meas}^X$ and $\lambda_{meas}^{ref}$, which correspond, respectively, to the measured LSD of a test sample and a reference object, one can compute the calibrated LSD value of the test sample $\lambda_{cal}^X$ as:

$$\lambda_{cal}^X = \lambda_{meas}^X \frac{\lambda_{pred}^{ref}}{\lambda_{meas}^{ref}} = \lambda_{meas}^X C_F \qquad (4.3)$$

where $\lambda_{pred}^{ref}$ is the predicted LSD value of the reference object and $C_F$ the calibration factor.

Four reference objects ("blocks") are situated near the four corners of the lower muon detector. All four blocks are parallelepipeds with surface area of $\{20 \times 20\}$ cm$^2$ in the horizontal plane. Two of the blocks, which are composed of iron, have a height of 10 cm, and the the other two blocks ("mock-up"), which have a custom structure, have a height of 10.5 cm. The two mock-up blocks are both composed of two smaller components, each of dimensions $\{(10 \times 20) \times 10.5\}$ cm$^3$, which are each formed of alternating layers of iron and plywood, with a density of $0.43 \pm 0.01$ kg/dm$^3$. The thicknesses of each of the iron and plywood layers are 2.0 mm and 8.5 mm, respectively, such that each component is formed of twenty layers (ten per material). The LSD of the two materials are $\lambda^{Fe} = 56.8$ rad$^2$/m and $\lambda^{plywood} = 1.1$ rad$^2$/m, such that, from Eq. 2.5, the LSD of the mock-up reference blocks is $\lambda^{mock-up} = 11.7$ rad$^2$/m. Since a better measurement precision is achieved if the LSD of the reference object is similar to the one of the sample object, only the mock-up blocks are used to calibrate the measurements of the EBF samples.

All data have been acquired over a period of approximately six months. Although the measurement system has proven to be particularly stable, the distribution of the LSD values of mock-up blocks, which is reconstructed from each data-taking "run", has been observed to be slightly wider than what is expected from the statistical uncertainty. Furthermore, a correlation is observed (within each run) between the LSD measurements of the mock-up and iron blocks. Whatever the reason for this effect, for example small variations of the muon momentum spectrum related to natural temporal variations of the muon flux [15], it is possible to minimise this variability by calibrating on a run-by-run basis. Hereafter, only calibrated $\lambda$ values are used.

### 4.6 Background subtraction

In order to account for the effects of the material containers, the internal structure of the muon tomography station, air and the effects of track measurement uncertainties, several runs are taken with empty buckets or without mock-up blocks. This allows one to subtract the LSD obtained for the empty bucket, and the volume in the vicinity of the mock-up blocks, from the corresponding values obtained with the objects in place. This subtraction is made prior to the calculation of $C_F$. The background subtraction process cancels some systematic uncertainties in the LSD extraction procedure.

### 4.7 Analysis of material samples

The bucket samples are analyzed by the following procedure. The bucket content is assumed to be homogeneous, since only the average bulk density is evaluated. Furthermore, small fluctuations in the material distribution are largely indistinguishable given the spatial resolution of the technology. Therefore, the value of LSD, $\lambda^{sample}$, assigned to a given sample material is calculated as:

$$\lambda^{sample} = \frac{V_{sample}}{V_{bucket}} \sum_i \lambda_i^{bucket} \quad (4.4)$$

where $\lambda_i^{bucket}$ is the LSD in voxel $i$ of volume $V_{bucket}$, and where $V_{bucket}$ is a volume that is sufficiently large as to fully encapsulate the bucket, and $V_{sample}$ is the effective volume of the sample material. Since the density reconstruction algorithm produces a non-negligible "halo" of LSD in the vicinity of the material, this procedure is necessary in order to avoid underestimating the average LSD. It is observed that the halo is more pronounced in the vertical direction, which reflects the fact that the majority of muons have trajectories close to the vertical direction. To obtain results independent of the choice of voxels, the same procedure is followed to evaluate the reconstructed LSD of the mock-up blocks, and consequently the calibration factor $C_F$.

An alternative procedure to determine the LSD of a sample is to exclude the boundary voxels from the analysis, for a depth compatible with the space resolution of the reconstructed image. However this will leave a small volume to work with, and hence the statistical error will increase widely and additional systematic uncertainties will have to be considered.

## 5 Results

This Section describes the measured values of LSD and of the quantity $R$, as defined in Section 2, for all samples considered in Section 3. Section 5.1 reports the results of the measurements of calibration and EBF materials.

For all measurements, the statistical uncertainty is obtained from the root mean square (r.m.s.) deviation of the results, with respect to the mean LSD calculated from the data subsets (described in Section 4.2). A discussion of all of the uncertainties affecting the measurements is provided in Section 5.2.

For all samples, results are presented by comparing the predicted LSD and $R$, $\lambda_{pred}$ and $R_{pred}$, with the experimentally measured values, $\lambda_{meas}$ and $R_{meas}$. The deviation of the measured value of $R$ with respect to the expectation, $\Delta R = (R_{meas}/R_{pred} - 1)$, is also calculated.

### 5.1 Measurements

The calibration material samples (described in Section 3) are measured in order to provide some quantitative validation of the LSD measurement procedure. The predicted values of $\lambda$ and $R$ can be computed through Eq. 2.5, once the chemical composition is known, with a precision better than 1% for all samples. The measured values of $R$ are expected to be valid for values of bulk density calculated using volumes which are sufficiently large compared to the size of the iron-ore pellets, which occupy a volume of approximately 1 cm$^3$.

The expected LSD for coke and for the iron oxides are separated by one order of magnitude, between 1.3 and 14 rad$^2$/m, and therefore the values of LSD for the EBF samples are expected to be within this range. The results which have been obtained for the calibration samples are presented in Table 2. As one could expect from Eq. 2.4, all oxides present similar values of LSD and $R$. A chemical analysis was performed to detect the presence of small amounts of additional elements in the calibration samples, which are taken into account in the calculation of the predicted LSD and $R$ values.

| Name | Material | $\lambda_{pred}$ (rad$^2$/m) | $\lambda_{meas}$ (rad$^2$/m) | $R_{pred}$ (rad$^2$m$^2$/ton) | $R_{meas}$ (rad$^2$m$^2$/ton) | $\Delta R$ (%) |
|---|---|---|---|---|---|---|
| A1 | Hematite | 13.65 | 12.74±1.08 | 5.84 | 5.45 ±0.37 | -7 |
| A2 | Magnetite | 12.55 | 11.28±0.95 | 5.92 | 5.32±0.36 | -10 |
| A3 | Wustite | 12.93 | 11.90±1.0 | 6.08 | 5.59±0.38 | -8 |
| A4 | Semi-metallic iron | 12.24 | 11.53±0.97 | 6.68 | 6.29±0.43 | -6 |
| A5 | Coke | 1.30 | 1.39±0.16 | 2.50 | 2.67±0.27 | 7 |
| A6 | Water (4.6 l) | 2.77 | 2.84±0.22 | 2.77 | 2.84±0.22 | 3 |
| A7/8 | Water (2.4 l) | 2.77 | 2.76±0.20 | 2.77 | 2.76±0.20 | 0 |

**Table 2**. A comparison of the predicted and measured values of LSD and $R$ for the calibration samples. $\Delta R$ is defined as $\Delta R = (R_{meas}/R_{pred} - 1)$.

By using the chemical composition of the probe and excavation materials, similar results are shown in Table 3 for the probe and excavation samples. The deviation $\Delta R$ is consistent with the expected uncertainty for these samples. The distribution of the measured LSD and $R$ values for all the samples, as a function of the effective bulk density, is shown in Figure 4 (left panel). The comparison of the measured and predicted values of $R$ is also shown (right panel).

### 5.2 Evaluation of uncertainties

The LSD measurements are affected by several sources of uncertainty, which are described in this Section.

- *Statistics.* The scattering of muons in a given material is a stochastic process. To obtain a stable evaluation of the material's properties, a large number of muons have to be considered. Once a dataset is sufficiently large, each volume unit is crossed by many muons and then the tomographic algorithm will return a stable value of LSD for that unit. As discussed in Section 4.2, LSD images are produced for each sub-run of 136 minutes. The LSD of any object is evaluated by averaging the results of all sub-runs. The statistical uncertainty on the average is obtained from the r.m.s. deviation of the results of the individual sub-runs. Given the large number of data, the statistical uncertainty for the analysis of bucket samples is smaller than 5%. Since there is a statistical uncertainty contribution due to the background subtraction, the materials with small LSD have the largest relative statistical uncertainty.

- *Algorithm parameters and convergence process.* The density reconstruction algorithm follows an iterative procedure requiring a large number of iterations to converge. Furthermore, the convergence may depend on additional parameters such as the initial values of LSD assigned to voxels, the object LSD and the size of datasets as discussed in Section 4. Here we have evaluated the global effect of the algorithm asymptotic convergence by changing the size of the subsamples. Varying the exposure time from 136 minutes to 460 minutes, the differences of LSD values have a global spread of 5%. This value is taken as an estimate of the systematic uncertainty due to the convergence process.

| Name | $\lambda_{pred}$ (rad$^2$/m) | $\lambda_{meas}$ (rad$^2$/m) | $R_{pred}$ (rad$^2$m$^2$/ton) | $R_{meas}$ (rad$^2$m$^2$/ton) | $\Delta R$ (%) |
|---|---|---|---|---|---|
| B1 | 3.71 | 3.74 ±0.32 | 4.22 | 4.26 ±0.30 | 1 |
| B2 | 4.79 | 4.83 ±0.43 | 4.65 | 4.69 ±0.34 | 1 |
| B3 | 9.83 | 9.46 ±0.80 | 5.61 | 5.40 ±0.37 | -4 |
| C1 | 5.62 | 5.53 ±0.48 | 5.19 | 5.10 ±0.36 | -2 |
| C2 | 3.94 | 3.85 ±0.33 | 4.55 | 4.44 ±0.31 | -2 |
| C3 | 6.52 | 6.63 ±0.57 | 5.54 | 5.63 ±0.39 | 2 |
| D1 | 4.34 | 4.30 ±0.38 | 4.70 | 4.65 ±0.34 | -1 |
| D2 | 9.18 | 8.68 ±0.75 | 6.14 | 5.80 ±0.41 | -5 |
| E1 | 2.06 | 2.08 ±0.21 | 3.42 | 3.44 ±0.30 | 1 |
| E2 | 6.11 | 6.28 ±0.54 | 5.08 | 5.22 ±0.37 | 3 |
| E3 | 11.27 | 10.56 ±0.89 | 5.59 | 5.24 ±0.36 | -6 |
| F1 | 2.75 | 2.78 ±0.26 | 3.83 | 3.86 ±0.31 | 1 |
| F2 | 5.00 | 4.90 ±0.43 | 4.84 | 4.75 ±0.34 | -2 |
| F3 | 9.15 | 9.06 ±0.77 | 5.54 | 5.48 ±0.38 | -1 |
| G1 | 2.01 | 1.99 ±0.21 | 3.31 | 3.28 ±0.31 | -1 |
| G2 | 7.42 | 7.65 ±0.66 | 5.58 | 5.75 ±0.40 | 3 |
| G3 | 8.66 | 8.34 ±0.70 | 5.64 | 5.34 ±0.37 | -4 |
| H1 | 1.79 | 1.80 ±0.19 | 3.21 | 3.22 ±0.30 | 0 |
| H2 | 3.27 | 3.49 ±0.32 | 4.25 | 4.53 ±0.34 | 7 |
| H3 | 6.38 | 5.86 ±0.51 | 5.31 | 4.88 ±0.34 | -8 |
| I1 | 3.73 | 3.37 ±0.32 | 4.43 | 4.00 ±0.32 | -10 |
| I2 | 7.25 | 7.05 ±0.60 | 5.75 | 5.60 ±0.39 | -3 |
| I3 | 9.40 | 8.50 ±0.73 | 5.94 | 5.37 ±0.37 | -10 |
| J1 | 2.90 | 2.93 ±0.27 | 3.76 | 3.80 ±0.30 | 1 |
| J2 | 7.44 | 6.83 ±0.60 | 5.87 | 5.39 ±0.39 | -8 |
| J3 | 7.94 | 7.80 ±0.68 | 5.77 | 5.67 ±0.40 | -2 |
| K1 | 2.75 | 2.55 ±0.25 | 3.96 | 3.67 ±0.30 | -7 |
| K2 | 8.00 | 7.16 ±0.61 | 6.41 | 5.73 ±0.40 | -11 |
| K3 | 7.05 | 6.60 ±0.56 | 6.09 | 5.69 ±0.39 | -6 |
| L1 | 5.26 | 4.94 ±0.44 | 5.25 | 4.93 ±0.37 | -6 |
| L2 | 7.67 | 6.69 ±0.58 | 6.05 | 5.27 ±0.37 | -13 |
| L3 | 8.74 | 8.01 ±0.68 | 6.27 | 5.74 ±0.39 | -8 |

**Table 3**. A comparison of the predicted and measured values of LSD and $R$ for the samples of EBF material. $\Delta R$ is defined as $\Delta R = (R_{meas}/R_{pred} - 1)$.

- *System stability.* As discussed in Section 4.4, the stability of the system is good but deviations are observed on like-for-like measurements between runs which are incompatible with statistical fluctuations. There may be several explanations for this residual instability, such as natural temporal variations in the cosmic ray average momentum [15], however as the effect is relatively small it is sufficient to recall that the calibration procedure is performed on a run-by-run basis, for which a residual uncertainty of 1.5% is calculated.

- *Uniformity.* The data are recorded with the relevant objects placed in different positions inside the inspection volume. Although the muon chambers ensure a good uniformity of response in terms of the measured track parameters, some residual effects may be present for several reasons. Such effects may be due (for example) to the presence of additional material, or to a muon-spectrum dependence on the position of objects due to the angular acceptance of the detectors, or to the presence of noisy channels. It has been observed that the background subtraction depends slightly on the position, but the variations are of the order of 0.1 rad$^2$/m. The variations of the measured LSD of like-for-like objects (mock-up blocks, iron blocks and buckets with 2.4 liters water) are of the same order, with a significance of about 2 standard deviations considering statistical uncertainties. Therefore a contribution of ±0.1 rad$^2$/m has been assumed for this uncertainty.

- *Choice of voxels.* As discussed in Section 4.7 the choice of voxels to be included for the LSD evaluation follows a procedure that minimizes the effects on final results. However, due to systematic differences in the geometry and positions of mock-up blocks and buckets, we estimate a residual uncertainty of approximately 4%. This uncertainty is correlated between all the material samples.

- *Geometric effects.* The effective volume of the material inside the buckets is directly used in the LSD computation. The uncertainty on the volume evaluation therefore affects the results. As discussed in Section 3, the uncertainty for the volume occupied by the material on the buckets is of approximately 5% and this uncertainty is therefore implied in the LSD measurements. On the other hand, since the same uncertainty enters the calculation of the effective bulk density, the contribution cancels out when the ratio $R$ is considered.

The contributions discussed above are assumed to be independent sources of uncertainty and are therefore combined in quadrature. The uncertainty on $\lambda$, that is correlated between all the samples, is approximately 4%. The total uncertainty ranges from 8% to 11% for materials with low and high LSD values, respectively. With regards to the $R$ measurements that are unaffected by the volume uncertainty, the total uncertainty ranges from 7% to 10%.

## 5.3 Interpretation of results

The range of values of LSD reported in Table 2, which describes the calibration materials, fully includes the range of values of LSD measured in the material samples extracted from the EBF, which are reported in Table 3. All of the LSD measurements were performed without prior knowledge of the chemical composition in order to eliminate any procedural bias. One would therefore expect that the precision of the LSD and $R$ measurements should be within the uncertainty presented in Section 5.2. This is confirmed by the comparison of the measured values of LSD and $R$ with the ones computed by taking into account the chemical composition of the EBF samples. If one considers all samples, the deviations $\Delta R$ present an r.m.s. of 4.9% with a mean value of -3.2%.

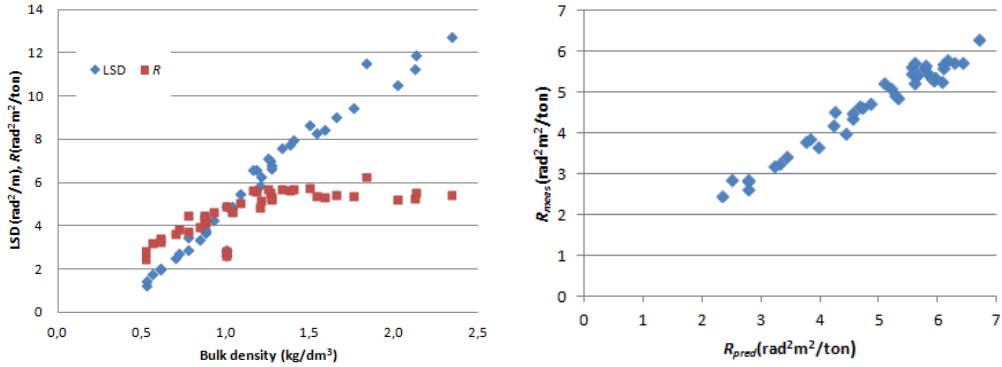

**Figure 4**. LSD and $R$ as a function of the effective bulk density for the calibration, probe and excavation samples (left panel). $R$ as a function of $R_{pred}$ for the same samples (right panel). The total uncertainty is listed in Tables 2 and 3 for all points.

It is observed that the measurements tend to overestimate the lowest values of LSD (the coke sample) while they underestimate the highest values of LSD (the iron compounds). In general, the deviation $\Delta R$ decreases as a function of the amount of material $x = L\rho$, where $L$ is an estimate of the average path length of particles inside the samples. This trend could be attributed to a non-linearity of the measurement when passing from light to heavy materials due to the absorption of the low momentum muons present in the cosmic spectrum [7]. By quantifying the effect one can extend the range of validity of our measurements to heavier materials. The values of $x$ for the measured samples range from $x = 7.8$ g/cm$^2$ for coke, to $x = 32.5$ g/cm$^2$ for hematite. We also consider two iron blocks with dimensions $\{(20 \times 20) \times 5\}$ cm$^3$ and $\{(20 \times 20) \times 10\}$ cm$^3$, respectively, with corresponding values $x = 39.3$ g/cm$^2$ and $x = 78.7$ g/cm$^2$. The fractional deviations $\Delta R$ are -11% and -22% in the two cases. The deviation decreases with increasing $x$, and for a value of $x$ which is more than double than that of the calibration material with higher bulk density, the deviation is approximately -22%.

As demonstrated in Ref. [7], the measurement of muon momentum would minimize the bias due to stopping muons. Unfortunately, as discussed above, we cannot perform such a measurement. On the other hand, a specific correction to the bias could be introduced on the basis of our measured data. However, in Tables 2 and 3, this correction is not applied since the discrepancy is consistent with the uncertainties discussed in Section 5.2 and the total amount of material does not vary greatly in the considered sample.

## 6 Conclusions

A large set of measurements has been produced with muon scattering tomography at the LNL laboratory, using data collected from several samples from the LKAB experimental blast furnace. The muon-tomographic technique has demonstrated the ability to reproduce the expected properties of all the samples. In particular the technique has been shown to be able to reproduce the properties of constituent materials of the samples and to be able to distinguish the light components as coke from the iron-oxides. The precision of the measurements of Linear Scattering Density relative to the mass density of all the samples is within 7 to 10%. The range of validity of the results is extended to heavier materials where effects due to the absorption of low momentum muons arise. As predicted, given the physical principles of the method, it is not possible to distinguish the degree of reduction if the density of iron atoms is not significantly altered. However, given the variation of densities as a function of position within the EBF, it is clear that muon tomography is able to provide information about the EBF content. One must consider, however, that there is significantly more material present in a full-scale blast furnace. This effect could be evaluated with simulated data, although such a study is beyond the scope of this paper. The study of this specific aspect is ongoing as a part of the project mentioned in Ref. [10].

## Acknowledgements

This work has been supported by the European Commission through the Research Fund for Coal and Steel, Grant RFSR-CT-2014-00027 for the Mu-Blast project.